\documentclass[draft]{agujournal2018}
\usepackage{apacite}
\usepackage{url} 
\usepackage{amsmath}
\usepackage{graphicx} 
\usepackage{bm} 
\usepackage{hyperref}

\draftfalse

\journalname{JGR: Solid Earth}

\begin{document}

%
%


\title{Gravity field modeling using space frequency signal transfer technique between satellites}

%
%




\authors{Ziyu Shen\affil{1}, Wen-Bin Shen\affil{2,3}, Xinyu Xu\affil{2}, Shuangxi Zhang\affil{2}}


\affiliation{1}{School of Resource, Environmental Science and Engineering, Hubei University of Science and Technology, Xianning, Hubei, China}
\affiliation{2}{Time and Frequency Geodesy Research Center, Department of Geophysics, School of Geodesy and Geomatics, Wuhan University}
\affiliation{3}{Key Lab of Surveying Eng. and Remote Sensing, Wuhan University}




\correspondingauthor {WenBin Shen}{wbshen@sgg.whu.edu.cn}
\correspondingauthor {Xinyu Xu}{xyxu@sgg.whu.edu.cn}



\begin{keypoints}
\item Low- and high-orbit satellites are connected by frequency links controlled by precise clocks.
\item Gravitational potential along the orbit of low-orbit satellite is determined.  
\item Optic-clock-based Earth's gravity field could be established. 
\end{keypoints}

%
%


\begin{abstract}
Here we provide an alternative approach to determine the Earth's external gravitational potential field based on low-orbit target satellite (TS), geostationary satellites (GS), and microwave signal links between them.
By emitting and receiving frequency signals controlled by precise clocks between TS and GS, we can determine the gravitational potential (GP) at the TS orbit. 
We set the TS with polar orbits, altitude of around 500 km above ground, and three evenly distributed GSs with equatorial orbits, altitudes of around 35000 km from the Earth's center. 
In this case, at any time the TS can be observed via frequency signal links by at least one GS. 
In this way we may determine a potential distribution over the TS-defined sphere (TDS), which is a sphere that best fits the TS' orbits. 
Then, based on the potential distribution over the TDS, an Earth's external gravitational field can be determined.
Simulation results show that the accuracy of the potential filed established based on 30-days observations can achieve decimeter level if optical atomic clocks with instability of $1\times 10^{-17}\tau^{-1/2}$ are available. The formulation proposed in this study may enrich the approachs for determining the Earth's external gravity field.

\end{abstract}

%
%

%


%
%
%
%

\section{Introduction}
\label{sec:intro}

The Earth's gravity field is a fundamental physical field of the Earth. 
Since gravity field has various and significant applications in many fields and branches, its determination is one of the main tasks in geodetic community.
If the density distribution of the Earth is given, one may determine the gravitational potential (GP) field both inside and outside the Earth (namely in whole space domaion) by Newtonian integral formula, and consequently the gravity field in whole space is determined by applying the gradient operator to the gravity potential (geopotential) field, where geopotential is the sum of GP and the centrifugal force potential generated by the Earth's rotation. 
However, since the Earth's density distribution (e.g. the preliminary reference Earth model, PREM) \citep{Dziewonski1981-kj} was poorly determined, the gravity field determined based on density distribution cannot satisfy the general application requirements. 
Fortunately, one can determine the external gravity field with successive accuracy requirement if some kind of distribution related to gravity (for instance the gravity distribution or gravity potential distribution) over the Earth's surface (boundary) is given \citep{Hofmann-Wellenhof2005-ur}. 
How to determine a gravity field (or equivalently gravity potential field) based on the given distribution values on the boundary (e.g. the Earth's surface) is referred to as the geodetic boundary value problem (GBVP). 
A generally accepted approach to solve this boundary problem is to use the spherical harmonic analysis, which can successfully express the external gravity field once a full coverage of the gravity (or geopotential) measurements over the Earth's surface or a surface (e.g. a surface defined by the orbits of a flying satellite) enclosing the whole solid Earth is provided.  
To overcome difficulties in practical measurements on the Earth's surface, especially in mountain areas and ocean areas, there appeared different satellite-based gravity measurement techniques, which have their own advantages especially in the aspect of full coverage over the Earth. 

\cite{Kaula1966-ke} proposed the method of establishing gravity model by observing the orbit perturbation of artificial satellites, and solve the coefficient of GP. 
Since the new satellite gravity missions appeared (e.g., the CHAMP mission \citep{Reigber2002-uf} launched on 2000, the GRACE twin satellite mission \citep{Tapley2004-tc} launched on 2002, the GOCE mission launched on 2009), scholars have paid extensive attention on the recovery of satellite gravity field and various methods have been proposed, such as orbital perturbation \citep{Hwang2001-jc}, harmonic analysis \citep{Reubelt2003-yb}, satellite accelerations \citep{Ditmar2004-si} and energy integral \citep{Jekeli1999-gr,Han2002-oe,Visser2003-xw}.
These satellite-based gravity measurement techniques have their own advantages especially in the aspect of full coverage over the Earth.

For example, the Gravity Field and Steady-State Ocean Circulation Explorer (GOCE) mission, which aims to make detailed measurements of Earth's gravity field, leading to discoveries about gravity field determination and ocean circulation investigations \citep{Drinkwater2003-os,Hirt2012-yg}.
GOCE satellite system flies in a near-polar orbit with an altitude of about 250 km above the ground, consisting of an on-board three-axis gravity gradiometer, GNSS receiver, satellite-to-satellite tracking and relevant equipments \citep{Bock2011-la,Hirt2012-yg}. 
Concerning the global gravity field determination aspect, the GOCE's mission may map gravity field features with 1 to 2 cm accuracy for geoid undulations and about 1 mgal for gravity, down to scales of about 100 km, or spherical harmonic degree about 200 \citep{Pail2011-qy,Hirt2012-yg}.
Since the GOCE satellite retired in 2013, the current satellite gravity mission on-going is the GRACE Follow-On \citep{Kornfeld2019-er}, which is a twin satellite system with the height of about 500 km.
These satellite-based gravity measurements have greatly improved our understanding of the Earth's gravity field \citep{Flechtner2016-jp,Kvas2019-vj,Pail2019-ao}.

In recent years, thanks to the quick development of time and frequency science, the optical-atomic clocks (OACs) with stability and accuracy better than  $1\times 10^{-18}$ level in several hours have been developed in laboratory environment \citep{Mehlstaubler2018-da,McGrew2018-og,Huang2019-ez,Oelker2019-wm}. Especially, portable and on-board-satellite clocks with ultra-high stability will be available in the near future \citep{Schiller2012-ps,Altschul2014-og,Hannig2019-yh}.
This provides potential realization in the near future to determine the GP differences between a satellite and a ground station using precise atomic-clock-related frequency signal links based upon general relativity theory (GRT) \citep{Einstein1915-tv}.
Suppose a satellite sends frequency signals and two receivers on different ground stations receive the signals, then the geopotential difference between the two stations can be determined by observing the frequency shift \citep{Shen2011-fm}.
However, how to practically and precisely extract the frequency shift signals caused by the geopotential difference between a ground station and a satellite is a challenging problem, due to the fact that Doppler effects, ionosphere and troposphere effects contaminate the observations seriously. 
In order to overcome these difficulties, recently a more precise formulation of the satellite frequency signal transfer (SFST) approach based on tri-frequency combination technique was established \citep{Shen2016-lc,Shen2017-kg}, which aims to determine the GP difference between a satellite and a ground station or between two satellites at an accuracy level of several centimeters if high-precise frequency signal links are established. 
To precisely compare the frequency signals, the relative stability of clocks should reach about $10^{-18}$ in several hours (corresponding to about 1 cm in height) for the practical applications of SFST method in geodesy. 



Based on the SFST technique, in this study we formulate an alternative approach to determine the Earth's external gravity field, which is completely different from the conventional ones.
The basic idea, which was put forward several years ago by our group \citep{Shen2017-xg} is that the GP along a low-orbit target satellite (TS) can be determined using frequency signal links between TS and geostationary satellites (GSs). 
In section \ref{sec:m} we briefly introduce the relativistic geodesy and SFST technique, by which the GP difference between the TS and GS could be determined. 
Then we formulate an approach to show how to determine a GP distribution over a TS-altitude defined sphere (TSS), which is bounded by the flights of the TS and defined as the sphere that can best fit the TS's orbits. 
Given the GP distribution over the TSS, a global gravity field (or Earth's gravity model) could be determined. 
In sections \ref{sec:exp} and  \ref{ssec:gpf}, we conducted simulation experiments, and results show that the proposed approach in this study is prospective. 
In section \ref{sec:con} we summarize the main results and discuss relevant potential issues.

\section{Method}
\label{sec:m}

\subsection{Gravity frequency shift}
\label{ssec:rg}

The GRT predicts that the frequency (or tick rate) of a clock is related to the geopotentials at the place  where the clock is located.
Specifically, suppose two clocks are located at different positions $P$ and $Q$ where the geopotential values are $W_P$ and $W_Q$ respectively, and accurate to $c^{-2}$, the frequencies $f_P$ and $f_Q$ of the two clocks satisfy the following equation \citep{Weinberg1972-le,Bjerhammar1985-bd}

\begin{equation}
    W_P - W_Q = \frac{f_P - f_Q}{f}\cdot c^2 + O(c^{-4}) ,
    \label{eq:bj}
\end{equation}
where $c$ is the speed ​​of light in vacuum, $f = (f_P + f_Q)/2$, $O(c^{-4})$ are high order terms which can be neglected in the case that the two stations are in the vicinity of Earth.
If the clock frequencies $f_P$ and $f_Q$ are precisely measured and compared, the geopotential difference $W_P - W_Q$ between $P$ and $Q$ can be derived.
The study of geodesy problems (such as geopotential determination) by the method of clock comparison is regarded as relativistic geodesy \citep{Flury2016-bw,Puetzfeld2019-sl}.

\begin{figure}[hbt]
  \centering
  \includegraphics[width=0.9\textwidth,keepaspectratio]{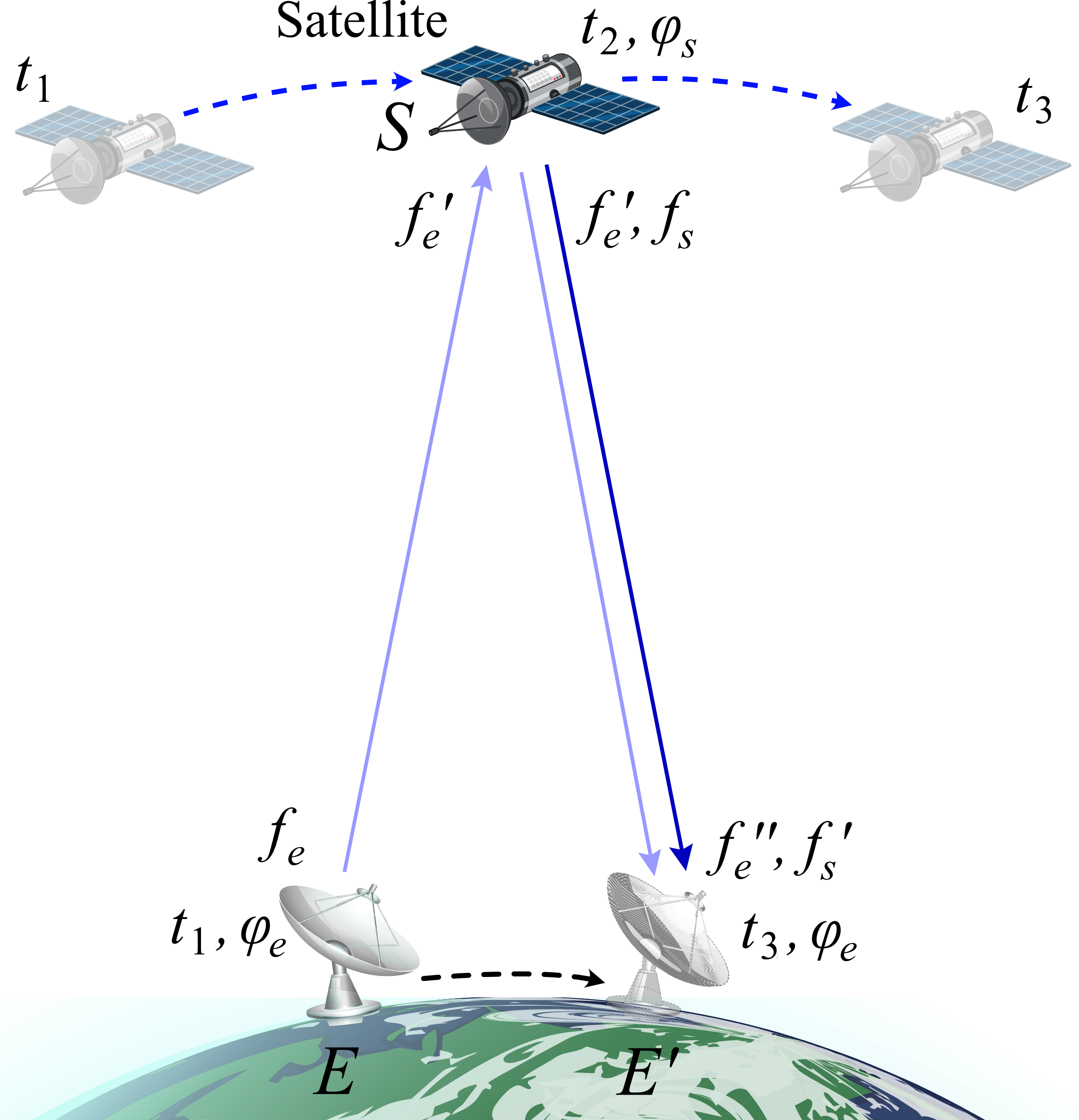}
  \caption{Ground station $E$ emits a frequency signal $f_e$ at time $t_1$, denoted by uplink (blue line). Satellite $S$ transmits the received signal $f_e'$ (the downlink denoted by blue line) and emits a new frequency signal $f_s$ at time $t_2$ (the downlink denoted by dark-blue line). The ground station receives signals $f_e''$ and $f_s'$ at time $t_3$ at position $E'$. $\phi$ is gravitational potential (GP).}
  \label{fig:oriobt}
\end{figure}

Currently, there are three kinds of method to compare clocks located at different places: (1) clock transportation \citep{Kopeikin2016-dk,Grotti2018-ap}, (2) transfer frequency signals via optical fibre links \citep{Takano2016-dr,Wu2019-wq,Shen2019-od}, and (3) transfer frequency signals via satellite and free-space links \citep{Deschenes2016-pr,Shen2017-kg}.
The first two methods are suitable for clocks comparison on ground, while the third method is designed for satellite clocks comparison.
But transferring frequency signals via satellite is much more complex than Eq. \eqref{eq:bj}. 
For example, the satellite is in high-speed motion state which gives rise to Doppler effects; the mediums in space (such as ionosphere and troposphere) will also cause frequency shifts during a microwave or optical signal's propagation through them.
In order to address these problem, \cite{Kleppner1970-db} proposed a method to transfer microwave frequency signal between a satellite and a ground site; and it is successfully applied for verifying Einstein's equivalence principle \citep{Vessot1979-ot,Vessot1980-ax}.
The main idea of the frequency transfer method is that a satellite and a ground site are connected by 3 microwave links simultaneously (as depicted in Fig. \ref{fig:oriobt}). In this case the first order Doppler effect and most of the medium influence will be canceled out in the output beat frequency $\Delta f$
\begin{equation}
  \frac{\Delta f}{f_e}  = \frac{f_s'-f_s}{f_e}-\frac{(f_e''-f_e')+(f_e'-f_e)}{2f_e} .
  \label{eq:outfrq}
\end{equation}
where $f_e$ and $f_s$ are emitted frequency from ground site and satellite respectively; they are received as $f_e'$ and $f_s'$ at satellite and ground site respectively. 
For each microwave link, the emitted frequency values are different from the received values. 
When the frequency signal $f_e'$ is received at satellite, it is transmitted immediately and received at ground site as $f_e''$.
Details can be referred to \citet{Vessot1979-ot}.

Kleppner's method was later improved and introduced to relativistic geodesy for GP determination \citep{Shen2016-lc,Shen2017-kg}, and was regarded as satellite frequency signal transmission (SFST) method.
According to SFST, the GP difference between a satellite and a ground site is given in the following form \citep{Shen2017-kg}
\begin{equation}
    \frac{\Delta \phi_{es}}{c^2} \equiv  \frac{\phi_s-\phi_{e}}{c^2} = \frac{\Delta f}{f_e} - \frac{v_s^2 - v_{e}^2}{2c^2} - \sum^4_{i=1} q^{(i)} + \Lambda f + O(c^{-5}) ,
    \label{eq:sg}
\end{equation}
where $\Delta \phi _{es}$ is the GP difference between the satellite and the ground station, $v_s$ and $v_e$ are velocities of satellite and ground site respectively, 
$q^{(i)}$ ($i=1,2,3,4$) are quantities related to the positions and velocities of the satellite and ground site, second Newtonian potential, vector potential, and third- and forth-order terms,  $\Lambda f$ is the correction terms for ionospheric, tropospheric and tidal effect, $O(c^{-5})$ denote high order terms that can be neglected.
 Details can be referred to \citet{Shen2017-kg}.

The theoretical precision of Eq. \eqref{eq:sg} is at the $10^{-19}$ level, much better than the original formula applied in GP-A experiment where its theoretical precision is limited to $10^{-15}$.
The precision of SFST method for determining GP is about several centimeters, provided that the stability of OACs can reach $10^{-18}$ level \citep{Shen2017-kg}.

\subsection{Determination of gravitational potential along the  target satellite orbit}
\label{ssec:leo}

Suppose the GP value of a ground station is given, then we can determine the GP values of a satellite by establishing SFST links between them.
If the GP distribution (GPD) over a TSS (e.g. GOCE-type or GRACE-type satellite) is determined, the Earth's external gravity field can be determined correspondingly (see Sect. \ref{ssec:gpf}).
However, since the orbit of a satellite for the purpose of determining the gravity field is relatively close to ground (e.g., the height of GRACE satellite is about 500 km), only a short arc length of the orbit is visible to a certain ground station.
If we want to determine the GPD over the TDS, hundreds of ground datum stations with given GP values are needed in order to guarantee that the satellite can connect to at least 1 ground station at any time \citep{Shen2018-jq}, which is impractical for the foreseeable future. 

\begin{figure}[hbt]
  \includegraphics[width=0.9\textwidth]{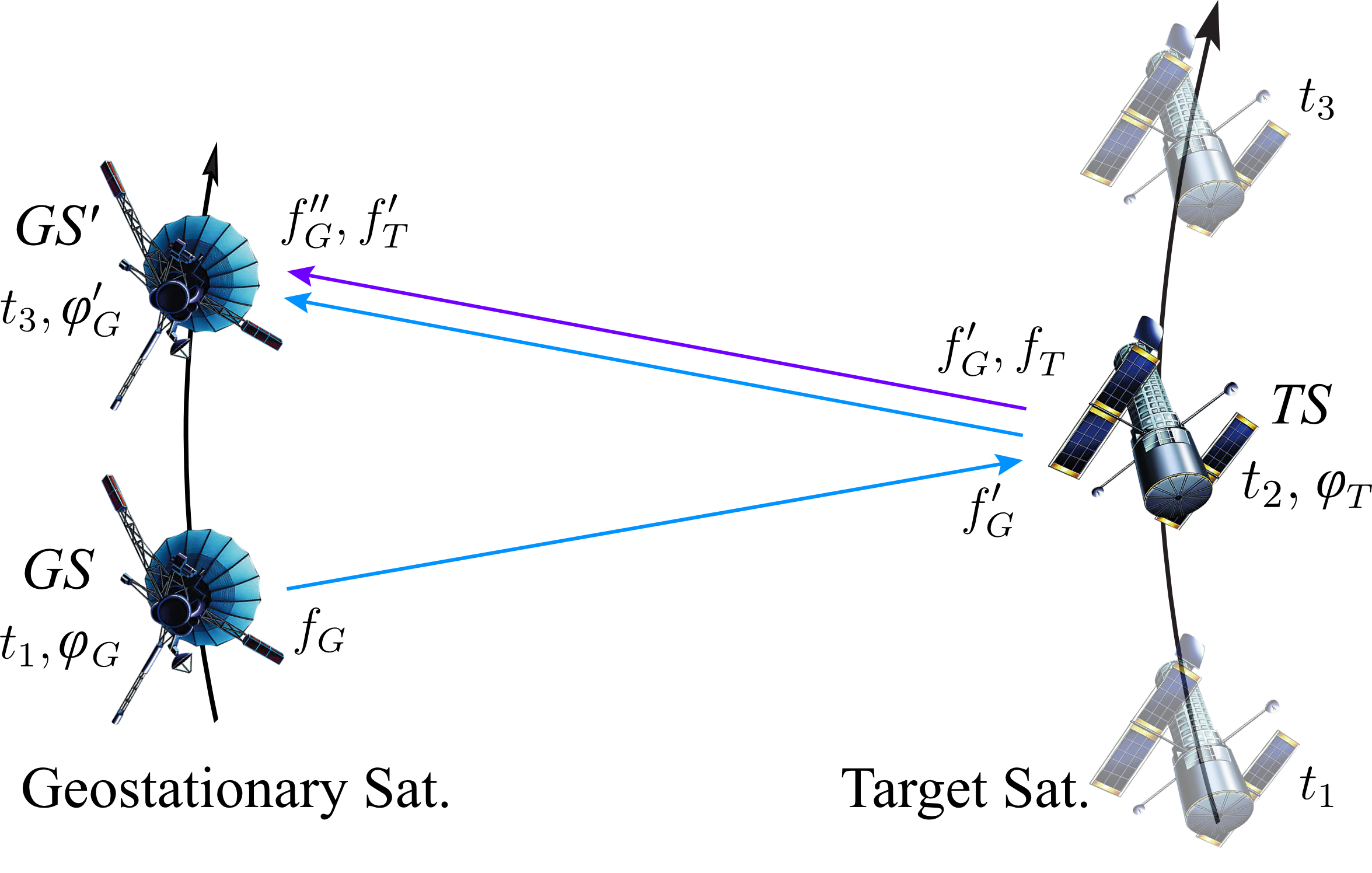}
  \caption{Geostationary satellite ($GS$) emits a frequency signal $f_G$ at time $t_1$, denoted by right arrow (blue line). Target satellite ($TS$) transmits the received signal $f_G'$ (the left arrow denoted by blue line) and emits a new frequency signal $f_T$ at time $t_2$ (the left arrow denoted by purple line). The geostationary satellite receives signals $f_G''$ and $f_T'$ at time $t_3$ at position $GS'$. $\phi$ denotes gravitational potential.}
  \label{fig:TG}
\end{figure}

Although the SFST method described in Sect. \ref{ssec:rg} was originally designed for determining the GP difference between a satellite and a ground site, it can also be used for determining the GP difference between two satellites after some modification \citep{Shen2019-hk}.
Suppose a TS (with low earth orbit) is connected to a GS (with high earth orbit), then the setup of the SFST links between them are depicted as Fig. \ref{fig:TG}. 

An emitter of the geostationary satellite $GS$ emits a frequency signal $f_G$ at time $t_1$. 
When the signal is received by the target satellite $TS$ at time $t_2$, it immediately transmits the received signal $f_G'$ and emits a frequency signal $f_T$ simultaneously. 
These two signals transmitted and emitted from the satellite are received by a receiver at geostationary satellite $GS$ at time $t_3$, which are noted as $f_G''$ and $f_T'$, respectively. 
During the period of the emitting and receiving, the position of the geostationary satellite in space has been changed from $GS$ to $GS'$.
Since the target satellite transmits and emits signals at the same instant it receives signal; its position in the signal links is supposed to be the point $TS$ at time $t_2$.
If we set $f_G = f_T$, the gravitational difference between the $GS$ and $TS$ can be expressed as
\begin{equation}
    \frac{\Delta \phi_{GT}}{c^2} \equiv  \frac{\phi_T-\phi_{G}}{c^2} = \frac{\Delta f}{f_G} - \frac{v_T^2 - v_{G}^2}{2c^2} - \sum^4_{i=1} q^{(i)} + \Lambda f + O(c^{-5}) ,
    \label{eq:GT}
\end{equation}
where the foot mark $G$ and $T$ denote the $GS$ and $TS$ respectively, and the beat frequency $\Delta f$ is given by.
\begin{equation}
  \frac{\Delta f}{f_G}  = \frac{f_T'-f_T}{f_G}-\frac{(f_G''-f_G')+(f_G'-f_G)}{2f_G} .
  \label{eq:ofrqGT}
\end{equation}

Note that there might be a small amount of latency during the transmitting, thereby the positions of the satellite is slightly different at the time it receives and emits signals. 
Suppose the delay of the signal transponder is about 800 ns \citep{Pierno2013-ub}, and the orbit height of TS is about 500 km (the velocity is about 7.6 km/s); then the satellite moves only 0.62 mm between receiving and emitting signals, and this influence can be neglected for the SFST links \citep{Shen2016-lc}.

\begin{figure}[hbt]
  \includegraphics[width=0.9\textwidth]{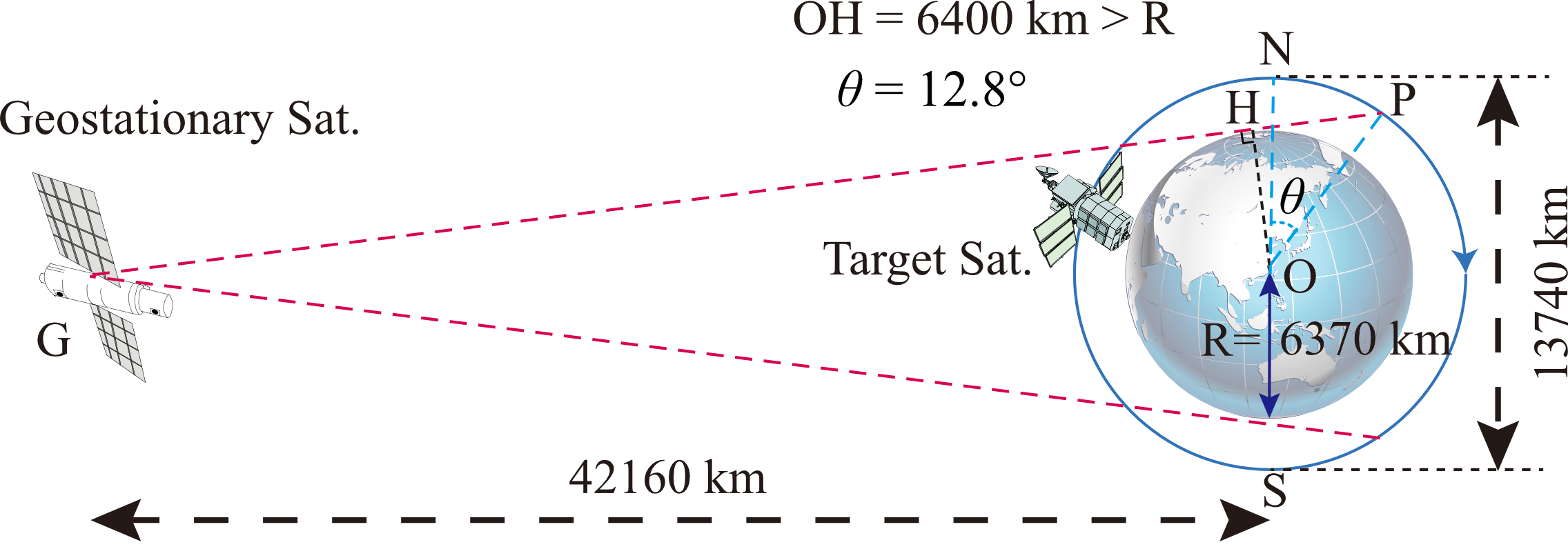}
  \caption{Suppose the height of TS is 500 km above the geoid and the Earth's  radius R is about 6370 km, then the distance between Earth's center and the target satellite is about 6870 km. The distance between Earth's center and a geostationary satellite is about 42170 km. When the TS is located at the poles (N or S), it can connect the GS without being blocked by the Earth, for the TS is visible until it reach the P point (if block threshold OH $=$ 6400 km, the angle of PON $\theta = 12.8^\circ$). Therefore the two satellites are inter-visible for more than half of the orbit period of the target satellite, and consequently two evenly distributed GSs are sufficient for incessantly SFST links between a GS and the TS.}
  \label{fig:size}
\end{figure}

Similar to the satellite-to-ground link, if the GP difference between the GS and TS are measured by SFST method, and the absolute GP values of the GS is given, then the GP values of the TS can be derived.
The height of a GS is about 35790 km above the equator.
If the TS is a low orbit satellite satellite whose height is about 500 km above the geoid  (e.g., the case of GRACE-FO satellite), then the GS can cover more than half of the TS's orbit sphere, as depicted in Fig. \ref{fig:size}.
Therefore only two evenly distributed GS is sufficient for incessantly SFST links between a GS and the TS, and the GP values of the TS's orbit sphere can be determined correspondingly. 
However, in practice it is better to apply three evenly distributed GSs for more reliable and stable connections, as shown in Fig. \ref{fig:3sat}

If the GPD over the TSS is given, one can derive the gravity field outside the TSS, and according to the spherical harmonic expansion formula, the determined gravity field outside the TSS can also be expanded to the Earth's surface \citep{Heiskanen1967-ue}.

\section{Simulation Experiments}
\label{sec:exp}

In this section we conducted several simulation experiments to verify the SFST method for satellite gravity model establishment.
Currently, the most precise atomic clock onboard a satellite is only about $10^{-13}\tau^{-1/2}$ ($\tau$ in second) in stability \citep{Laurent2015-ly,Liu2018-oo}.
While the best optical atomic clock on ground have reached the stability of $4.8\times 10^{-17}\tau^{-1/2}$ \citep{Oelker2019-wm}.
In the prospect of much better clocks onboard satellites in the future, our experiments will adopt different clock stability levels from $10^{-13}\tau^{-1/2}$ to $10^{-17}\tau^{-1/2}$; and the results can show us the minimum requirements of clock stability for a satellite gravity model in a certain precision.

\begin{figure}[hbt]
  \centering
  \includegraphics[width=0.9\textwidth,keepaspectratio]{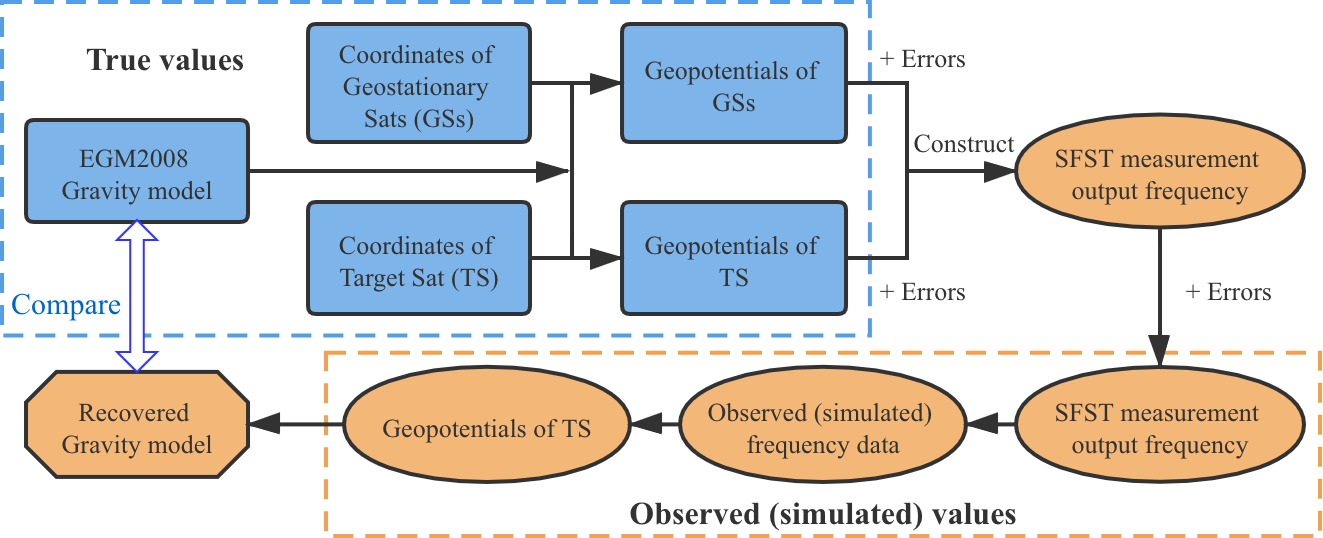}
  \caption{The scheme of the simulation experiment.}
  \label{fig:scheme}
\end{figure}

The scheme of a simulation experiment is comparing a prior satellite gravity model to a recovered one, as depicted in Fig. \ref{fig:scheme}; details are explained in the following subsections.

\subsection{Input data}
\label{ssec:in}

\begin{figure}[hbt]
  \centering
  \includegraphics[width=0.9\textwidth]{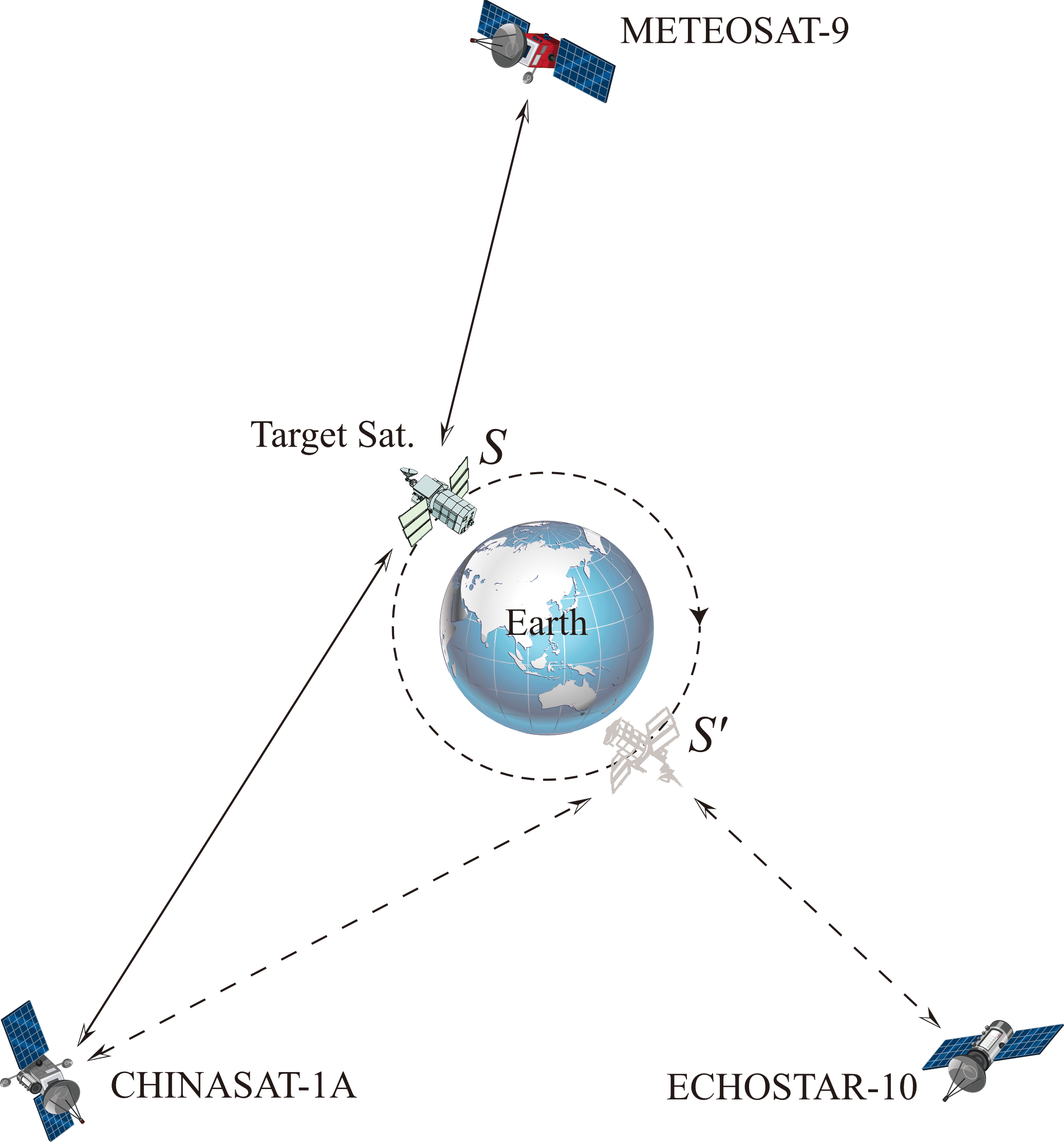}
  \caption{The TS (target satellite) is connected with 3 GSs (geostationary satellites). When the TS flies around the Earth, it is continuously connected with one or two GSs via SFST links. Given the gravitational potentials at orbits of GSs, the gravitational potential at the orbit of TS can be obtained.}
  \label{fig:3sat}
\end{figure}

In Sect. \ref{ssec:leo} we have shown that two GSs are sufficient for incessant SFST links to a TS.
But in practice it is more reliable to adopt three evenly distributed GSs above the equator.
Therefore in our experiments we chose the meteorological satellite METEOSAT-9 of EU (at 9.2$^\circ$E), the communication satellite CHINASAT-1A of China (at 130.0$^\circ$E), and the communication satellite ECHOSTAR-10 of US (at 110.2$^\circ$W) as the GSs; and the GRACE-FO 1 satellite (orbit height is about 500 km) as the TS.
The setup of the experiment is depicted in Fig \ref{fig:3sat}. 

The orbital period of TS (GRACE-FO 1) is about 1.6 h, and the inclination of TS is about 89$^\circ$.
For the purpose of obtaining the GP data over its orbit sphere with a resolution of $30'\times 30'$, the observations should continue for at least 576 hours (24.0 day), with observation interval being smaller than 8.0 second.
Therefore, we set the observational time span as 30.0 days, and the observation interval as 1 second, which fully satisfies the resolution requirements.
For every observation the TS is connected to a nearest visible GS with SFST links.
For the purpose of simulation, the orbits data of these four satellites (one TS and three GSs) can be generated from two-line element set (TLE) data by simplified general perturbations models 4 (SGP4) \citep{Romania2016-aa}. 
The GPs at orbits of these satellites can be calculated by EGM2008 model \citep{Pavlis2012-fw}.
Then the GP difference between the TS and one proper GS can be obtained.
These data are all regarded as true values, hence the errors of orbit data and gravitational potential model EGM2008 are not considered.

The frequency of a microwave signal will be affected by ionosphere and troposphere environment. 
We adopt the International Reference Ionosphere Model \citep{Rawer1978-fl,Bilitza2017-zr} to obtain the electron density values to estimate the ionospheric influence \citep{Namazov1975-qa}.
Since the height of TS is about 500 km which is much higher than the troposphere layer (typically from ground to 60 km height), the influence of troposphere can be neglected.
The GP at the satellites' orbits will also be influenced by periodical tidal effects, which are well modeled \citep{Voigt2017-qc} and can be removed by some mature softwares such as ETERNA \citep{Wenzel1996-tk} or Tsoft \citep{Van_Camp2005-qt}.
In our experiment we use ETERNA to generate and analyze tide signals. These tidal signals also include the influences of other planets (such as Venus, Jupiter etc.) besides the Sun and the Moon.


\begin{table}
    \caption{The input datas used in simulation experiments.}
    \label{tab:in}
    \footnotesize
    \begin{tabular}{@{}ll}
        \hline\noalign{\smallskip}
        Entities & Values of Parameters \\
        \noalign{\smallskip}\hline\noalign{\smallskip}
        GS Satellite & METEOSAT-9 \\
        & CHINASAT-1A \\
        & ECHOSTAR-10 \\
        TS Satellite & GRACE FO 1\\ 
        Gravity field model & EGM2008 \\
        Ionospheric model & International Reference Ionosphere \\
        Tide correction & ETERNA \\
        Observation duration & Jan 01 $\sim$ Jan 30, 2020 \\
        Mearsurement interval & 1 s \\
    \noalign{\smallskip}\hline\noalign{\smallskip}
    \end{tabular}
\end{table}

\subsection{The ``Observed'' GP along the TS orbit}
\label{ssec:obs}

After setting the input data, the next step focuses on determining the GP values at TS's orbit.
There are 3 GSs, denoted as $GS_i~(i=1,2,3)$ respectively.
As we take a SFST measurement (every 1 second), we can obtain an observed GP difference value $\Delta \hat \phi_{GTi}(t)$ according to Eq. \eqref{eq:GT}.
If the GP of $GS_i(t)$ is given, the observed GP $\hat \phi_{T}(t)$ can be derived as $\hat \phi_{T}(t) = GS_i(t) - \Delta \hat \phi_{GTi}(t)$.

The observed values $\hat \phi_{T}(t)$ are different from true GP value $\phi_{T}(t)$ because they are influenced by various error sources.
In this simulation experiment we have considered clock error $e_{clk}$, ionosphere residual error $e_{ion}$, satellite's position and velocity errors $e_{pos}$ and $e_{vel}$, GS's potential errors $e_{pot}$ and tidal correction residual error $e_{tide}$.
The above mentioned various errors are considered as noises, which are added to the true values. 
The total errors $e_{all}$ are expressed in the following form
\begin{equation}
    e_{all} = e_{clk} + e_{ion} + e_{pos} + e_{vel} + e_{pot} + e_{tide} ,
    \label{eq:e-all}
\end{equation}
and the observed values $\hat \phi_{T}(t)$ can be expressed as
\begin{equation}
    \frac{\hat \phi_T(t)}{c^2} = \frac{\phi_{G}(t)}{c^2} + \frac{\Delta f(t)}{f_G} - \frac{v_T(t)^2 - v_{G}(t)^2}{2c^2} - \sum^4_{i=1} q^{(i)} + \Lambda f(t) +e_{all}(t) ,
    \label{eq:GT_exp}
\end{equation}
The magnitude and behavior of each kind of error play important role in this experiment; thereby we need to investigate different error models based on different error sources to make the simulation case more close to the real case.

\begin{figure}[hbt]
  \centering
  \includegraphics[width=0.9\textwidth,keepaspectratio]{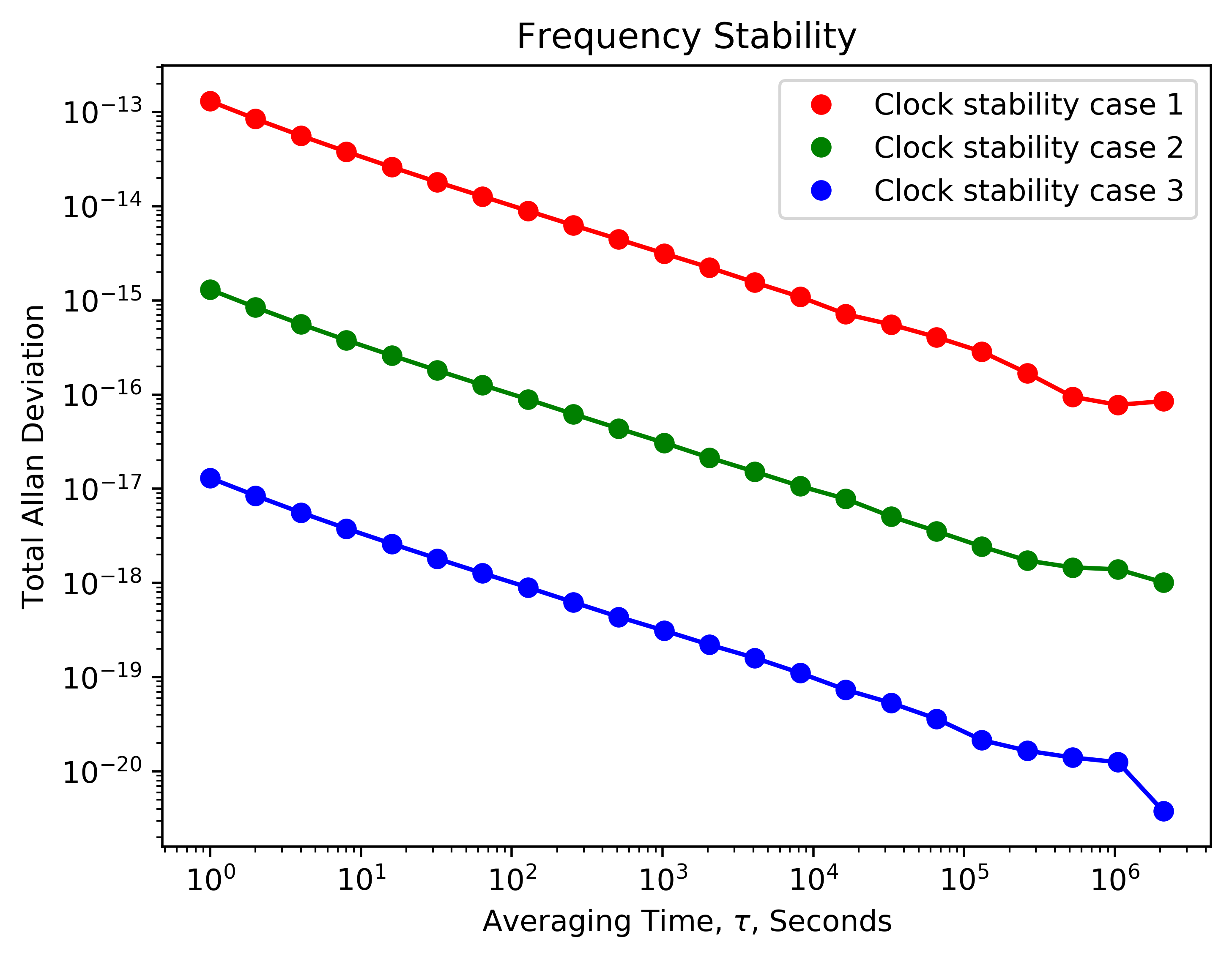}
  \caption{The total Allan deviation for three different clocks. The instabilities of the clocks are about $10^{-13}\tau^{-1/2}$, $10^{-15}\tau^{-1/2}$ and $10^{-17}\tau^{-1/2}$ for case 1, case 2 and case 3 respectively.}
  \label{fig:allan}
\end{figure}

We first set the clock error magnitude of $e_{clk}$ as $1.0\times 10^{−13}\tau^{-1/2}$, which is achievable currently.
Considering the present best clocks with stability of $1\times 10^{-18}$ in several hours, we reduce the magnitude of $e_{clk}$ to $1.0\times 10^{-15}\tau^{-1/2}$ and $1.0\times 10^{-17}\tau^{-1/2}$ respectively to improve the observations. 
Although there are many kinds of random noises that affect Atomic clocks' signals \citep{Major2013-qc}, the most prominent components are white frequency modulation and random walk frequency modulation \citep{Galleani2003-hu}.
Correspondingly the behaviors of clock errors are modeled as following equation
\begin{equation}
    e_{clk}(t) = a_{clk} + b_{clk}\cdot t + c_{clk}\cdot \phi (t) + d_{clk}\cdot \int_0^t \xi (t)dt ,
    \label{eq:e-clk}
\end{equation}
where $a_{clk}$, $b_{clk}$, $c_{clk}$ and $d_{clk}$ are constant coefficients, $\phi (t)$ and $\xi (t)$ are both standard white Gaussian noises.
Each term in the right side of Eq.\eqref{eq:e-clk} has clear physical meaning; specifically $a_{clk}$ denotes the initial frequency difference, $b_{clk}\cdot t$ is the drift term, $c_{clk}\cdot \phi (t)$ is the white noise component, and $d_{clk}\cdot \int_0^t \xi (t)dt$ represents the random walk effect.
As we set proper values of constant coefficients in accordance with the performance of OACs in \citet{Oelker2019-wm}, a series of frequency comparison data with errors embedded can be generated.
The statistic property of three clock error series are shown in Fig. \ref{fig:allan}.

Other error sources are discussed in detail in \citet{Shen2017-kg}.
Although \citet{Shen2017-kg} focus on the satellite to ground case, we can use similar methods to analyze the errors in the satellite to satellite case, and most of them demonstrate the same magnitude.
The magnitudes of these error sources are listed in Table \ref{tab:err}.
 
As for the mathematical model of these errors, we adopt a general error model which contains systematic (initial) offset, drift and white Gaussian noises for each of the error source, expressed as the following equation
\begin{equation}
    e_j(t) = a_j + b_j\cdot t + c_j\cdot \phi_i (t) ,~~~~~(j=ion, tro, pos, vel, tide, asy)
    \label{eq:e-other}
\end{equation}
where $a_j$, $b_j$ and $c_j$ are constant coefficients, which are randomly set in accordance with the error magnitudes listed in {Table} \ref{tab:err}

\begin{table}
    \caption{Error magnitudes of different error sources in determining the gravitational potential difference between a satellite and a ground station. They are transformed to relative frequency. Details can be referred to \citet{Shen2017-kg}.}
    \footnotesize
    \begin{tabular}{@{}lll}
      \hline\noalign{\smallskip} 
      Influence factor & (Residual) Error magnitude in $\Delta f / f_e$ \\
      \noalign{\smallskip}\hline\noalign{\smallskip}  \\
      ionospheric correction residual & $\delta f_{ion} \sim 10^{-18}$  \\
      tidal correction residual & $\delta f_{tide} \sim 10^{-18}$ \\
      position \& velocity & $\delta f_{vepo} \sim 10^{-17}$ (10 mm and 0.1mm/s $^a$) \\
      clock error & $\delta f_{osc} \sim {10^{-13}\tau^{-1/2}}$ \\
      \noalign{\smallskip}\hline\noalign{\smallskip}  \\
    \end{tabular}
    \label{tab:err}
  $^a$ Satellite's position errors are assumed as 10 mm \citep{Kang2006-ix}, velocity errors are assumed as 0.1mm/s \citep{Sharifi2013-bm};
\end{table}

According to Eqs. \eqref{eq:e-clk} and \eqref{eq:e-other}, we can generate the noise signals $e_{all}(t)$ term in Eq. \eqref{eq:GT_exp} based on the magnitudes and nature of the error sources at any time. 
Noted that the first 4 terms or the right side of Eq. \eqref{eq:GT_exp} are true values, the 5th terms ($\Lambda f(t)$) is the corrections of ionosphere and tide effect.
The values of these 5 terms can be directly calculated.
Therefore we can get a set of relevant  "Observed" values, which constitute time series of TS's gravitational potential $\hat \phi_{T}(t)$ as the left side of Eq. \eqref{eq:GT_exp}.
Since the TS flies over the whole Earth in a period of about 30 d, these values are corresponding to the gravitational potential at different time points on the TS's orbits and we obtain a set of values $\hat \phi_{T}(x, y, z)$ related to orbit data.

\subsection{Determination of the GPD over the TSS} 
\label{ssec:res}
\begin{figure}
	\centering
	\includegraphics[width=1.0\textwidth,keepaspectratio]{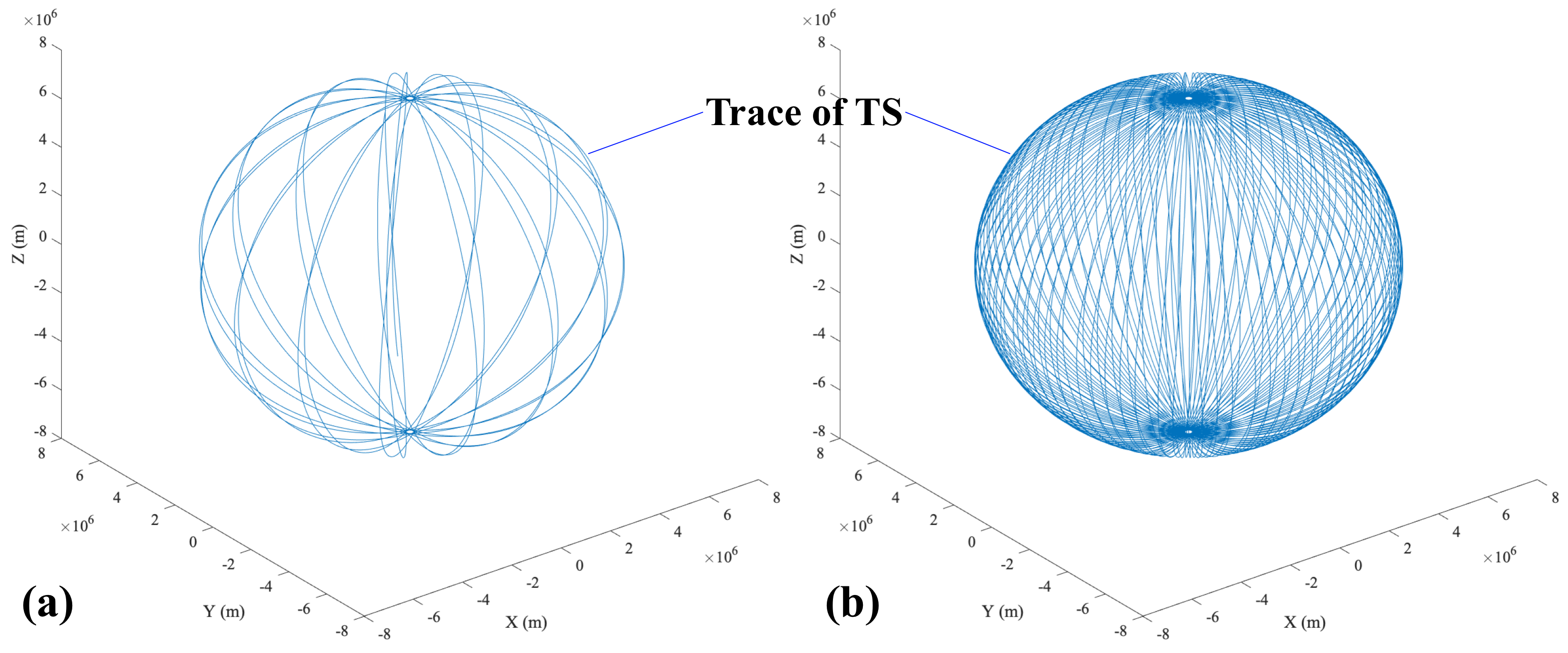}
	\caption{The trace of target satellite (TS) in Earth-Centered and Earth-Fixed (ECEF) coordinate for \textbf{(a)} 1 day and \textbf{(b)} 5 days.}
	\label{fig:trace}
\end{figure}

\begin{figure}[hbt]
  \centering
  \includegraphics[width=1.0\textwidth,keepaspectratio]{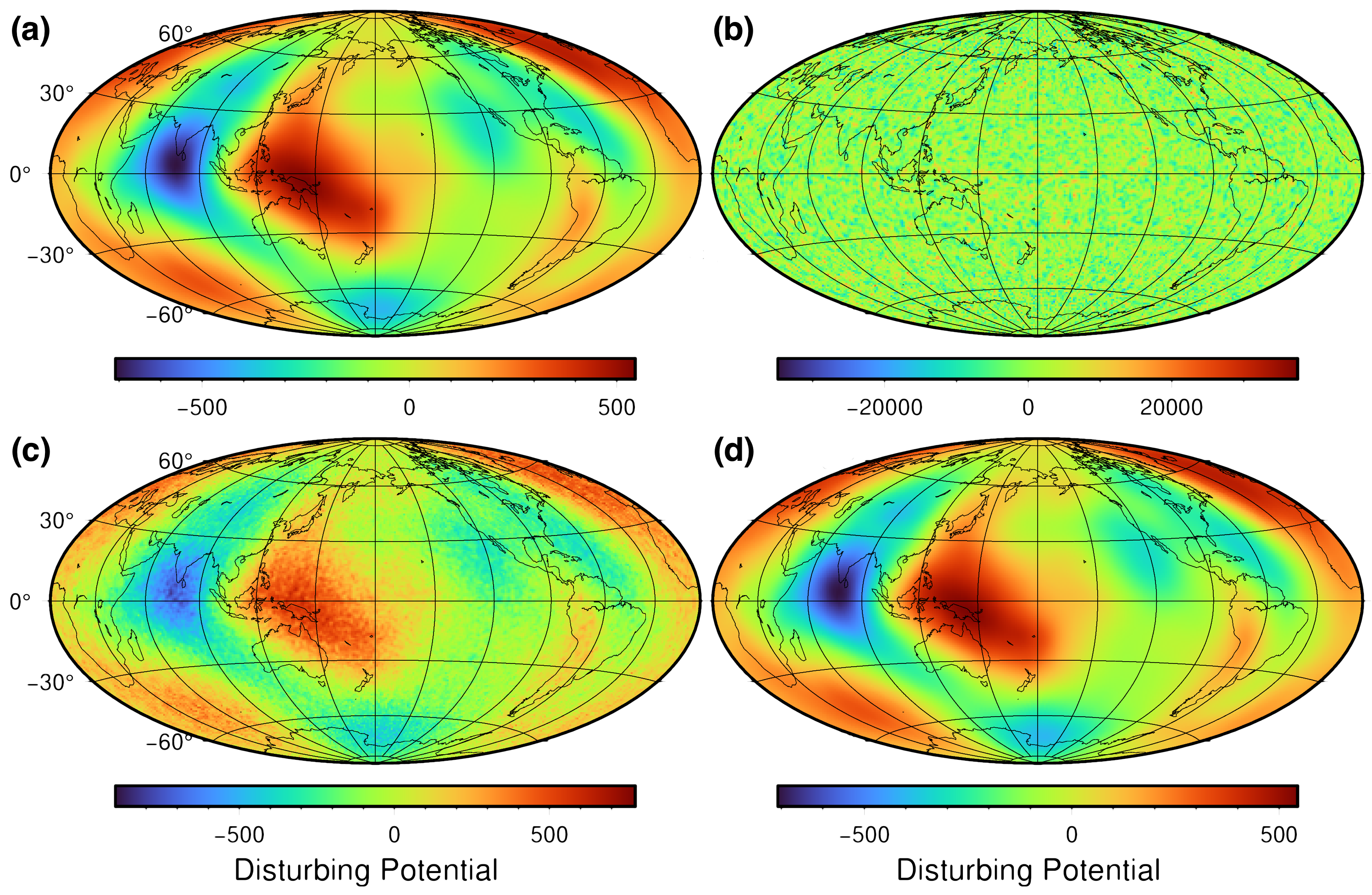}
  \caption{The disturbing potentials (m$^2$s$^{-2}$) of \textbf{(a)} EGM2008 as true values; \textbf{(b)} experiment case 1, using clocks with instability of about $10^{-13}\tau^{-1/2}$; \textbf{(c)} experiment case 2, using clocks with instability of about $10^{-15}\tau^{-1/2}$; \textbf{(d)} experiment case 3, using clocks with instability of about $10^{-17}\tau^{-1/2}$.}
  \label{fig:potres}
\end{figure}

After a continuous observation of 30.0 days (720.0 hours), there are 324,000 observed GP points $\hat \phi_{T}(x, y, z)$ distributed over the TSS enclosing the Earth (see Fig. \ref{fig:trace}).
The corresponding disturbing potentials of the three different experiment cases are depicted in Fig. \ref{fig:potres}, where the first subfigure demonstrates the disturbing potential of EGM2008 as true values.
we can see that the observe results of case 1 (clock instability of $10^{-13}\tau^{-1/2}$) seems to be useless, but the observe results of case 3 (clock instability of $10^{-17}\tau^{-1/2}$) are almost identical with the true values.
In the next section, we will calculate the spherical harmonic expansion coefficients for 3 different recovered Earths gravity field models (REGMs) based on the observed GP values in our simulation experiments.
These coefficients will be compared with the true value of EGM2008 to evaluate their accuracy.

\section{Determination of the Earth's external gravitational potential field}
\label{ssec:gpf}

Based on the determined  GPD over the TDS using frequency links, we can determine the gravitational potential field outside the solid Earth by least-squares (LS) approach. 
The Earth’s gravitational potential $V$ at a point $(r,\theta,\lambda)$ outside the Earth can be expanded into a series of spherical harmonics \citep{Hofmann-Wellenhof2005-ur}
\begin{equation}
  V(r,\theta,\lambda) = \frac{GM}{a}\sum_{n=0}^{N_{max}}\sum_{m=0}^n\left(\frac{a}{r}\right)^{n+1}\left(\overline C_{nm}\cos m\lambda + \overline S_{nm}\sin m\lambda\right)\overline{P}_{nm}\left(\cos \theta\right),
  \label{eq:sh}
\end{equation}
where the spherical coordinates $(r,\theta,\lambda)$ represent a 3-D position in the Earth-Centered, Earth-Fixed (ECEF) reference frame, $r$ is the geocentric radius, $\theta$  and $\lambda$  are the spherical co-latitude and longitude respectively, $GM$ is the geocentric gravitational constant, $a$ is the semi-major axis of the reference ellipsoid, $\overline C_{nm}$ and $\overline S_{nm}$ the (fully-normalized) geopotential coefficients which describe the external gravitational field of the Earth, $\overline P_{nm}$ are the (fully-normalized) associated Legendre functions of degree $n$ and order $m$, and $N_{max}$ is the maximum degree of the harmonic expansion. 

For the linear observation equation Eq. \eqref{eq:sh}, the functional and statistical models of the gravitational field recovery from the GPD observations are defined by a standard Gauss-Markov model as follows:
\begin{equation}
  \bm{y} = \bm{A}\bm{x} + \bm{\epsilon},\quad E\left\{\bm{y}\right\} = \bm{A}\bm{x},\quad D\left\{\bm{y}\right\} = \sigma_0^2\bm{Q} = \sigma_0^2\bm{P}^{-1},
  \label{eq:mkov}
\end{equation}
where $\bm{y}$  is the vector of GP observations, $\bm{A}$ is the design matrix, $\bm{x}$ is the vector of (unknown) geopotential coefficients $\overline C_{nm}, \overline S_{nm}$  to be estimated, $\bm{\epsilon}$  is the vector of observation errors, $D\left\{\bm{y}\right\}$ is the error variance-covariance matrix, $\bm{P}$ is the weight matrix, $\bm{Q}$ is the inverse of the weight matrix, and $\sigma_0^2$ is the variance component.

\begin{figure}[htbp]
  \centering
  \includegraphics[width=1.0\textwidth,keepaspectratio]{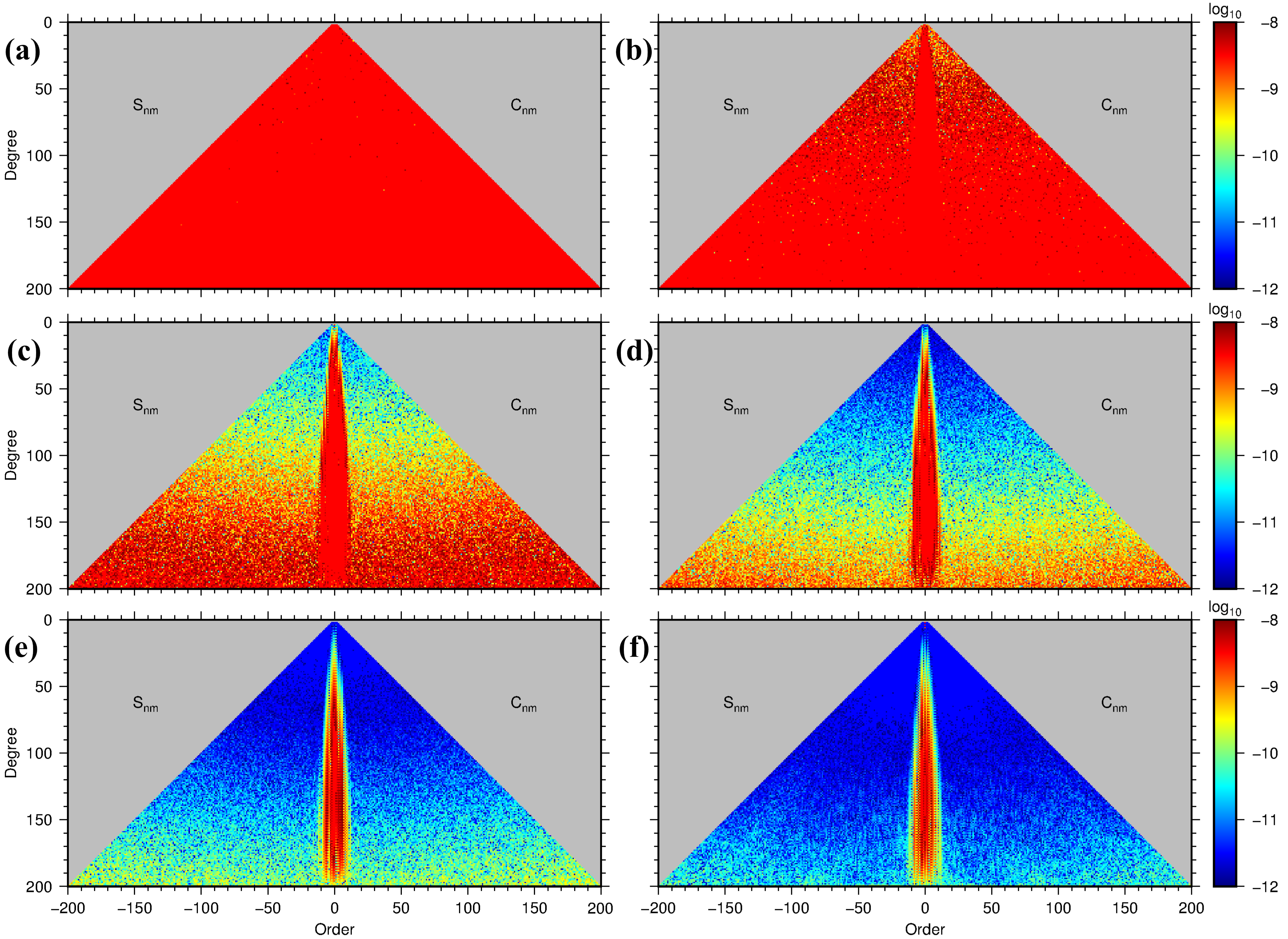}
  \caption{The offset between recovered harmonic expansion coefficients and that of EGM2008. The clock instabilities of different experiments are about \textbf{(a)} $10^{-13}\tau^{-1/2}$; \textbf{(b)} $10^{-15}\tau^{-1/2}$; \textbf{(c)} $10^{-17}\tau^{-1/2}$. \textbf{(d)} and \textbf{(e)} are another two demonstration experiments which we set the total error magnitude for a SFST link is about $10^{-18}$ and $10^{-19}$ respectively. \textbf{(f)} denotes the recovered coefficients based on true values of GPD over the TDS, determined by EGM2008}
  \label{fig:cp357}
\end{figure}

Based on the data processing method described above, we estimated three REGMs up to degree and order 200 from GP values distributed over the TSS in three different cases, corresponding respectively the clock’s instabilities of $10^{-13}\tau^{-1/2}$, $10^{-15}\tau^{1/2}$ and $10^{-17}\tau^{-1/2}$. 
Here we set the weight matrix $\bm{P}$ as a unit matrix by considering that the noise in GP observations is white noise. The absolute values of the coefficient differences (logarithm representation) between the recovered harmonic expansion coefficients of Earth’s gravity field and that of EGM2008 are illustrated globally in Fig. \ref{fig:cp357} as (a), (b) and (c).

The results show that if the clock instability is poorer than $10^{-17}\tau^{-1/2}$ (as cases (a) and (b)),the precision of recovered Earth's gravity fields is poor.
When the clock instability reach $10^{-17}\tau^{-1/2}$ as case (c), we can obtain a fairly good coefficients for orders and degrees lower than 50.
In addition, the zonal and near-zonal coefficients of REGMs are worse than other kinds of coefficients when the clock’s instability reaches $10^{-15}\tau^{-1/2}$. 
This is due to the fact that we did not use any regularized technique to deal with the ill-posed problem caused by the polar gap of GOCE mission. 
However, even if we use the regularized technique to deal with the ill-posed problem, the above mentioned problem still exists \citep{Baur2014-oa}.

The performance of REGMs can also be evaluated by the calculated GPs at the TSS (TS-orbit defined spherical surface).
The mean offset and standard deviation (STD) of the calculated GP distribution difference between EGM2008 and REGMs over TSS are illustrated in Table \ref{tab:res}.
The STDs of GPD over the TSS are respectively 1135.3232 $\rm{m^2/s^2}$, 11.3758 $\rm{m^2/s^2}$ and 0.114 $\rm{m^2/s^2}$ for cases (a), (b) and (c).

\begin{table}
  \caption{The statistic information of the GP offset at target satellite's orbit between values calculated by EGM2008 (true values) and by recovered Earths gravity field models (REGMs).}
  \label{tab:res}
  \footnotesize
  \begin{tabular}{@{}llll}
      \hline\noalign{\smallskip}
      Case & Clock precision & Mean offset ($\rm{m}^2/\rm{s}^{2}$) & STD ($\rm{m}^2/\rm{s}^{2}$) \\
      \noalign{\smallskip}\hline\noalign{\smallskip}
      (a) & $10^{-13}\tau^{-1/2}$ & 0.9142 & 1135.3232 \\
      (b) & $10^{-15}\tau^{-1/2}$ & 0.2094 & 11.3758 \\
      (c) & $10^{-17}\tau^{-1/2}$ & -0.0009 & 0.1142 \\
      (d) & $10^{-18}\tau^{-1/2}$ & -0.0001 & 0.0114 \\
      (e) & $10^{-19}\tau^{-1/2}$ & -2.5e-6 &  0.0013 \\
      (f) & NA & 2.4e-8 &  0.0006 \\
  \noalign{\smallskip}\hline\noalign{\smallskip}
  \end{tabular}
\end{table}

If the clock instabilities can be improved to $10^{-18}\tau^{-1/2}$ or even $10^{-19}\tau^{-1/2}$ level, some other error sources, such as ionospheric residual and velocity error, will take dominant and become the bottle neck for the precision of recovered coefficients.
In that case, further detailed analysis or corrections for various error sources is required for establishing better correction models.
Since there might be an extended period for us to set onboard clocks of $10^{-18}\tau^{-1/2}$ instability level, in this paper we will leave that researches for future works.
However, in order to show the potential of this method, we conducted two simplified experiments that
The total error (the sum of clock errors and various other error sources) of SFST links are set to $10^{-18}$ and $10^{-19}$ respectively as illustrated in Fig. \ref{fig:cp357} (d) and (e).
We can see that if the total error magnitudes can reduced to $10^{-19}$, the recovered harmonic expansion coefficients show fairly good quality, close to that recovered from the true values as illustrated in Fig. \ref{fig:cp357} (f).
The REGM's precision of case (d), (e) and (f) are shown in table \ref{tab:res}.

\section{Conclusion}
\label{sec:con}

In this paper we formulated an alternative method to determine satellite gravity field based on precise clocks and frequency signals transfer.
It is a new application of general relativistic theory in geodesy, and the gravity field can be determined at the precision levels of about $10^3~\rm{m^2/s^2}$, $10~\rm{m^2/s^2}$ and $10^{-1}~\rm{m^2/s^2}$, given the clock stabilities of $10^{-13}\tau^{-1/2}$, $10^{-15}\tau^{-1/2}$ and $10^{-17}\tau^{-1/2}$ respectively.
Currently the stability of a satellite's onboard clock is about $10^{-13}\tau^{-1/2}$, and it is the main error influence for GP determination and EGM establishment.
However, precise optical atomic clocks have reached $10^{-17}\tau^{-1/2}$ level under laboratory environments \citep{Oelker2019-wm}. 
It is foreseeable that in the near future the stability of onboard atomic clocks can achieve a similar level, and the precision of the EGM established by intersatellite SFST method can reach 1 cm level. 
Compared to the conventionally used methods of establishing satellite gravity model such as using gravimeter and gravity gradiometer to  measure the first-order and second-order derivative of potential, the SFST method may directly determine the GPs, simplifying in some sense the estimation of the harmonic coefficients of Earth's gravity field.

According to this study, once the onboard clocks' stabilities reach the level of $10^{-17}\tau^{-1/2}$, the relativistic method will be applicable for high precision satellite gravity field determination, and can be used to provide a decimeter level Earth gravity model with a resolution of around $1^\circ \times 1^\circ $.
If clock stabilities are better than $10^{-17}\tau^{-1/2}$ ($10^{-18}\tau^{-1/2}$ or even $10^{-19}\tau^{-1/2}$ for instance), various other error sources (ionosphere and troposphere correction residual errors, satellite position errors, et al.) will be the bottle-neck for determining a precise Earth's gravity field, and more precise error correction models need to be established. 

\acknowledgments
This study is supported by National Natural Science Foundation of China (NSFC) (grant Nos. 41721003, 41631072, 41874023, 41804012, 41429401, 41574007), and Natural Science Foundation of Hubei Province (grant No. 2019CFB611).


%
%

\bibliography{main}




\end{document}